\documentclass[journal,transmag]{IEEEtran}

\usepackage[export]{adjustbox}
\usepackage{graphicx}
\usepackage{array,tabularx}
\usepackage{multirow,booktabs} 
\usepackage{pifont}
\usepackage{tabularx,booktabs}
\usepackage{lipsum}
\usepackage{booktabs,threeparttable}
\usepackage{subfig}
\usepackage{mwe}

\usepackage[utf8]{inputenc} 
\usepackage[T1]{fontenc}    

\usepackage[colorlinks=true, allcolors=blue]{hyperref}
\newcommand{\email}[1]{\href{mailto:#1}{\tt{\nolinkurl{#1}}}}
\newcommand{\orcid}[1]{ORCID: \href{https://orcid.org/#1}{\tt{\nolinkurl{#1}}}}

\usepackage{url}            
\usepackage{booktabs}       
\usepackage{amsfonts}       
\usepackage{nicefrac}       
\usepackage{microtype}      
\usepackage{lipsum}
\usepackage{amssymb}
\usepackage{latexsym}

\usepackage{url}
\usepackage{xcolor}
\usepackage{hyperref}

\usepackage{amsmath}
\usepackage{graphicx}
\usepackage{cleveref}
\usepackage[export]{adjustbox}
\usepackage{subfig}
\usepackage{mwe}
\usepackage{graphicx}
\usepackage{pdflscape}

\usepackage[english]{babel}
\usepackage[utf8]{inputenc}
\usepackage{fancyhdr}
\usepackage{atbegshi}
\usepackage{lipsum} 

\begin{document}
%
\title{ROCT-Net: A new ensemble deep convolutional model with improved spatial resolution learning for detecting common diseases from retinal OCT images}


\author{\IEEEauthorblockN{Mohammad Rahimzadeh\IEEEauthorrefmark{1,*},
Mahmoud Reza Mohammadi\IEEEauthorrefmark{1,}\IEEEauthorrefmark{2}}
\IEEEauthorblockA{\IEEEauthorrefmark{1}FaraAI Company,
Tehran, Iran.}
\IEEEauthorblockA{\IEEEauthorrefmark{2}School of Electrical and Computer Engineering, University of Tehran, Tehran, Iran.}
\IEEEauthorblockA{\IEEEauthorrefmark{*} Corresponding author : \email{mh_rahimzadeh@faraai.ir}}
\thanks{978-1-6654-0208-8/21/\$31.00 ©2021 IEEE}}



\IEEEtitleabstractindextext{%
\begin{abstract}
Optical coherence tomography (OCT) imaging is a well-known technology for visualizing retinal layers and helps ophthalmologists to detect possible diseases. 
Accurate and early diagnosis of common retinal diseases can prevent the patients from suffering critical damages to their vision. Computer-aided diagnosis (CAD) systems can significantly assist ophthalmologists in improving their examinations. This paper presents a new enhanced deep ensemble convolutional neural network for detecting retinal diseases from OCT images. Our model generates rich and multi-resolution features by employing the learning architectures of two robust convolutional models. 
Spatial resolution is a critical factor in medical images, especially the OCT images that contain tiny essential points. To empower our model, we apply a new post-architecture model to our ensemble model for enhancing spatial resolution learning without increasing computational costs. The introduced post-architecture model can be deployed to any feature extraction model to improve the utilization of the feature map's spatial values.
 We have collected two open-source datasets for our experiments to make our models capable of detecting six crucial retinal diseases: Age-related Macular Degeneration (AMD), Central Serous Retinopathy (CSR), Diabetic Retinopathy (DR), Choroidal Neovascularization (CNV), Diabetic Macular Edema (DME), and Drusen alongside the normal cases.
Our experiments on two datasets and comparing our model with some other well-known deep convolutional neural networks have proven that our architecture can increase the classification accuracy up to 5\%. 
We hope that our proposed methods create the next step of CAD systems development and help future researches. The code of this paper is shared at \href{https://github.com/mr7495/OCT-classification}{https://github.com/mr7495/OCT-classification}.
\end{abstract}

}

\maketitle

\IEEEdisplaynontitleabstractindextext

%
\IEEEpeerreviewmaketitle

\section{Introduction}
\label{1}
\IEEEPARstart{O}ptical coherence tomography (OCT) is a medical imaging technology that captures cross-sectional image slices from the retinal structure \cite{GHOLAMI2020106532}. OCT images have been widely used for visualizing the biologic tissues \cite{rasti2017macular}.
This technology has made many advances in the early diagnosis of crucial retinal diseases that can affect the patient's retina and vision.

There are some major pathologies that can threaten a patient's sight and even lead to blindness, such as Age-related Macular Degeneration (AMD), Central Serous Retinopathy (CSR), Diabetic Retinopathy (DR), Choroidal Neovascularization (CNV), and Diabetic Macular Edema (DME). AMD has been announced as the fourth most common cause of blindness in 2013 \cite{vos2015global}.
Based on these statements, early detection of the described diseases is essential for saving the patient's retina. 

Computer-aided diagnosis (CAD) systems have been progressively utilized in the last years to assist ophthalmologists in improving their diagnosis quality and speed. As the manual examination of the 3D slices of OCT images can be hard and time-consuming, using the CAD system can enhance the analysis and diagnosis process.

Since the progress of deep learning and convolutional neural networks \cite{lecun1995convolutional}, computer vision tasks have been improved significantly. Currently computer vision applications play great roles in agriculture \cite{rahimzadeh2021detecting}, industry \cite{akinosho2020deep}, and especially medical diagnosis field \cite{rahimzadeh2021fully}.

One of the main challenges in computer vision is medical image classification. Since 2012 and the introduction of Alexnet \cite{krizhevsky2012imagenet}, convolutional neural networks prove to be enhancing feature extraction notably. After AlexNet \cite{krizhevsky2012imagenet}, VGG \cite{simonyan2015deep}, ResNet \cite{he2015deep}, DenseNet \cite{huang2018densely}, Inception \cite{szegedy2016rethinking}, NasNet \cite{zoph2018learning} models have been introduced and each achieved better results compared to its previous model. 

In 2019, a new family of convolutional neural networks called EfficientNet \cite{tan2020efficientnet} has been introduced, which raised the classification accuracy while reducing the number of model weights. In 2021, the authors of the same paper released an updated version of this family named EfficientNetV2 \cite{tan2021efficientnetv2}. Based on the authors' knowledge, this new family owns the best classification performance on the ImageNet dataset \cite{5206848} until now.

In this research paper, we have developed a new ensemble model created by concatenating the architectures of the EfficientNetV2-B0 \cite{tan2021efficientnetv2} and Xception \cite{chollet2017xception} models. Both of these models are robust and capable of generating rich semantic features with different scales. So by merging the features of these two models, we have been able to improve the learning and classification accuracy as the produced ensemble model can learn different kinds of features.

 The merged models have been chosen; somehow that the final ensemble model does not include many parameters and be easy for training and inference. We have also introduced a new post-architecture model for being applied to the ensemble model to advance the spatial resolution learning procedure. Spatial resolution is very considerable in medical images as the critical areas may be tiny, and the classic classification methods like adopting Global Average Pooling \cite{lin2013network}, or Fully-Connected layers for features compression may lead to losing important data or underfitting.

Now we mention some of the developed CAD systems. This paper \cite{rasti2017macular} proposed a CAD system based on a multi-scale convolutional mixture of expert (MCME) ensemble models to detect AMD, DME, and normal cases. \cite{wang2019oct} trained and evaluated several deep convolutional models using transfer learning to identify AMD and DME.

Authors of {schlegl2018fully} presented a fully automated system for diagnosing AMD, RVO, and DME with a mean accuracy of 0.94.
Another research \cite{lee2017deep} developed a deep convolutional model for classifying the OCT images into Normal and AMD and achieved 87.63\% accuracy.

The rest of this paper is organized as follows: In section \ref{2}, our methods and models have been expressed. Section \ref{3} illustrates the experimental results and comparison to other works, and in sections \ref{4}, and \ref{5} the paper is discussed and concluded, respectively.

\section{Material and Methods}
\label{2}

\subsection{Dataset}
\label{21}

Most of the OCT datasets include few images of 2 or 3 types of disease. In order to expand our investigations and validate our results, we ran our experiments on two datasets. The first is a large dataset \cite{kermany2018identifying} \footnote{ This dataset is available at https://data.mendeley.com/datasets/rscbjbr9sj/3}.
containing 108,312 images (37,206 with CNV, 11,349 with DME, 8,617 with DRUSEN, and 51,140 Normal) from 4,686 patients for training and 1000 images from 633 patients for evaluating the models.

The second dataset is named OCTID \cite{GHOLAMI2020106532} and belongs to the waterloo university. It includes 572 images of 5 classes: Normal, AMD, CSR, DR, and MH.

\begin{figure*}
\centering
\includegraphics[scale=0.144]{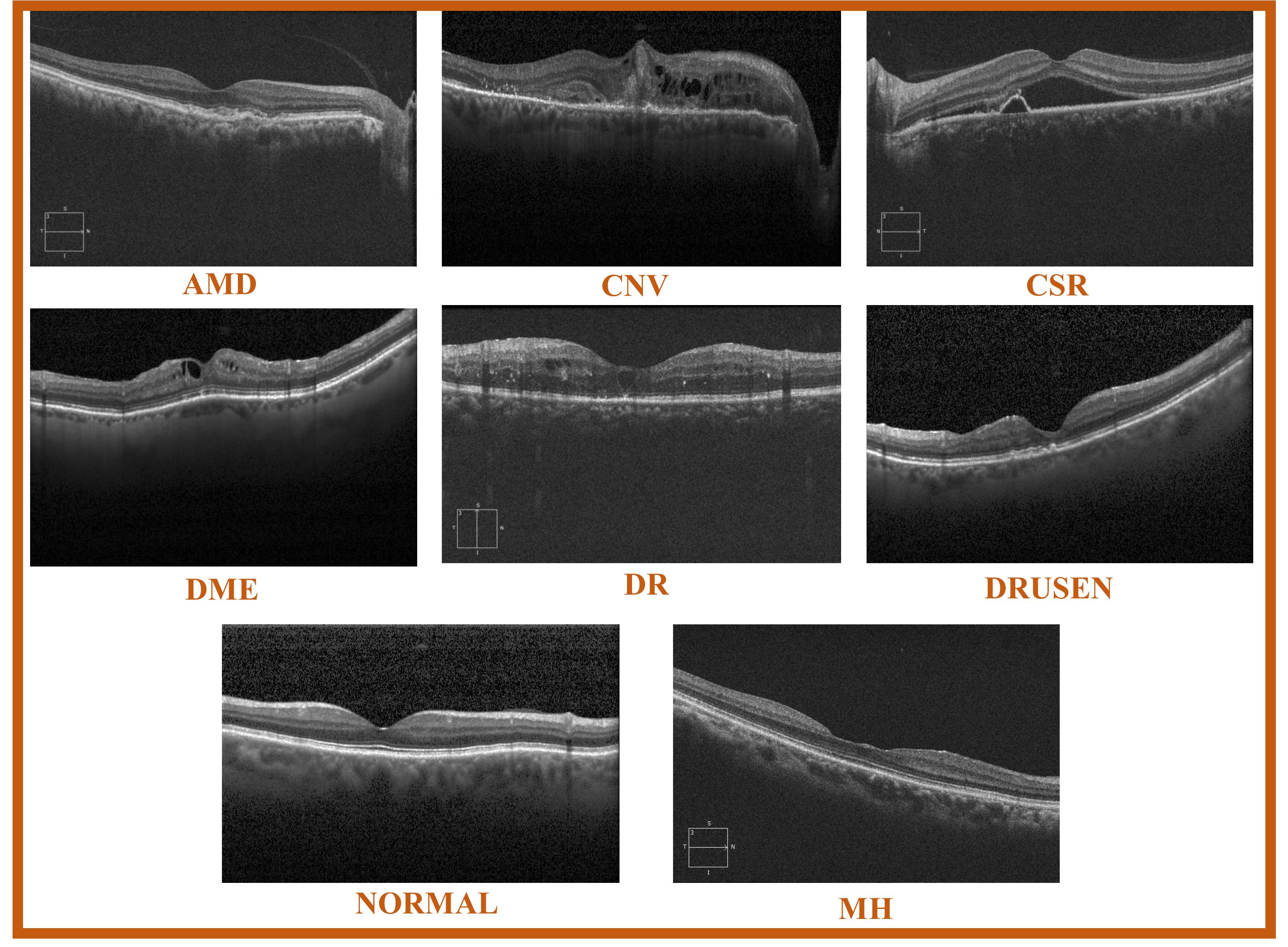}
\caption{One image of each disease class in our collected dataset.}
\label{images}
\end{figure*}

In the Kermany dataset \cite{kermany2018identifying}, each patient has several OCT images. We selected one image of each patient in the training set to reduce the data size, but we kept the same test set for evaluation. In the second dataset, we allocated 20\% of images of each class for testing, and the rest were used for training the models. The details of the data distribution of our work are presented in the table. \ref{dataset}. Some of the images in our collected dataset are displayed in fig. \ref{images}.

\begin{table*}[!ht]
\caption{This table shows the details of the utilized dataset in this paper.}
\label{dataset}
\centering
\large
\begin{adjustbox}{width=1\linewidth}
\begin{tabular}{lll|lcccccccc|}
\cline{4-12}
                              &                                         &                                                           &                                                                               & \multicolumn{1}{l}{}         & \multicolumn{1}{l}{}         & \multicolumn{1}{l}{}         & \multicolumn{1}{l}{}         & Train / Test Images            & \multicolumn{1}{l}{}           & \multicolumn{1}{l}{}           & \multicolumn{1}{l|}{} \\ \hline
\multicolumn{1}{|l|}{Dataset} & \multicolumn{1}{l|}{Number of  Classes} & \begin{tabular}[c]{@{}l@{}}Training\\ Images\end{tabular} & \multicolumn{1}{l|}{\begin{tabular}[c]{@{}l@{}}Testing\\ Images\end{tabular}} & \multicolumn{1}{c|}{AMD}     & \multicolumn{1}{c|}{CSR}     & \multicolumn{1}{c|}{DR}      & \multicolumn{1}{c|}{MH}      & \multicolumn{1}{c|}{CNV}       & \multicolumn{1}{c|}{DME}       & \multicolumn{1}{c|}{DRUSEN}    & NORMAL                \\ \hline
\multicolumn{1}{|l|}{Kermany} & \multicolumn{1}{l|}{4}                  & 3213                                                      & \multicolumn{1}{l|}{1000}                                                     & \multicolumn{1}{c|}{N/A}     & \multicolumn{1}{c|}{N/A}     & \multicolumn{1}{c|}{N/A}     & \multicolumn{1}{c|}{N/A}     & \multicolumn{1}{c|}{791 / 250} & \multicolumn{1}{c|}{709 / 250} & \multicolumn{1}{c|}{713 / 250} & 1000 / 250            \\ \hline
\multicolumn{1}{|l|}{OCTID}   & \multicolumn{1}{l|}{5}                  & 459                                                       & \multicolumn{1}{l|}{113}                                                      & \multicolumn{1}{c|}{44 / 11} & \multicolumn{1}{c|}{82 / 20} & \multicolumn{1}{c|}{86 / 21} & \multicolumn{1}{c|}{85 / 20} & \multicolumn{1}{c|}{N/A}       & \multicolumn{1}{c|}{N/A}       & \multicolumn{1}{c|}{N/A}       & 165 / 41              \\ \hline
\end{tabular}
\end{adjustbox}
\end{table*}

\subsection{Neural Networks}
\label{22}

\subsubsection{EfficientNetV2 and Xception}
\label{221}

Convolutional neural networks have been impressively developing in the last decade. Now, most of the computer vision models are build based on convolutional layers. In this paper, as we explore a classification problem, we seek to find the model that can generate more valuable features from the OCT images and obtains the best possible accuracy.

Since 2012 and the introdcution of AlexNet \cite{krizhevsky2012imagenet}, many families of convolutiolan neural network such as ResNet \cite{he2015deep}, DenseNet \cite{huang2018densely}, Inception \cite{szegedy2016rethinking}, NasNet \cite{zoph2018learning}, and MobileNet \cite{howard2017mobilenets} have been propsed.

Xception \cite{chollet2017xception} is a powerful convolutional model that introduced the depthwise separable convolutional layer for the first time. This layer is a compressed and updated version of the classic convolutional layer, containing much fewer parameters. Depthwise separable convolutional layer is made of a depthwise convolutional layer and pointwise convolutional layer to process the spatial and channel data. As this layer includes fewer parameters, utilizing it lets the developers expand the model and increase the number of layers. Xception was placed among the top classifiers in the ImageNet benchmark.

EfficientNet family \cite{tan2020efficientnet} was created using reinforcement learning and neural architecture search (NAS). The basic architecture of these models was inspired by the building blocks of MobileNet \cite{howard2017mobilenets} (MBConv layers) and then, running the NAS, scaled the width, depth, and resolution of its layers. It comes with eight models of different sizes (from EfficientNetB0 to EfficientNetB7). The evaluation results of this family show its superiority over other families.

In 2021, the second version of this family named EfficientNetV2 \cite{tan2021efficientnetv2} has been proposed that adopts a combination of training-aware and scaling methods and is made of an updated building block called Fused-MBConv. One of the best novelties of this version is the usage of progressive learning techniques to adaptively indicate training parameters like dropout value, data augmentation values, and image size during training. Utilizing the two approaches increased the training speed and efficiency somehow that let the authors train their models on ImageNet21K, which is a huge dataset and includes 21000 classes (ImageNet
ILSVRC2012 contains1000 classes!). The models of this family achieved perfect and interesting results.

\subsubsection{Ensemble Convolutional Model}
\label{222}

Ensemble learning is a well-known technique in deep learning. In this method, researchers combine the architecture of several models to improve learning efficiency. Implementing ensemble learning depends on the fact the constructor models be adaptable together, resulting in outputting higher semantic features. Combining incompatible models can damage the learning procedure and decrease performance.

In this phase of our work, based on the authors' 
experience and experiments, we decided to concatenate EfficientNetV2-B0 \cite{tan2021efficientnetv2} and Xception \cite{chollet2017xception} models. EfficientNetV2-B0 is from the latest generation of convolutional neural networks and achieves great results while having very few parameters (less than 6M params). 

On the other hand, Xception is also a well-performing model, and our investigation in various fields proves that it has a high learning capability in challenging situations. 

Xcpetion generates 2048 channels at its final produced feature map, while EfficientNetV2-B0 generates 1280 channels of data at the final layer. As the EfficientNetV2-B0  is a low-size model, the final ensemble model created from the concatenation of Xception and EfficienNetV2-B0 will not have many parameters (weights) and will be optimized for training and inference. 

\begin{figure*}
\centering
\includegraphics[width=\linewidth]{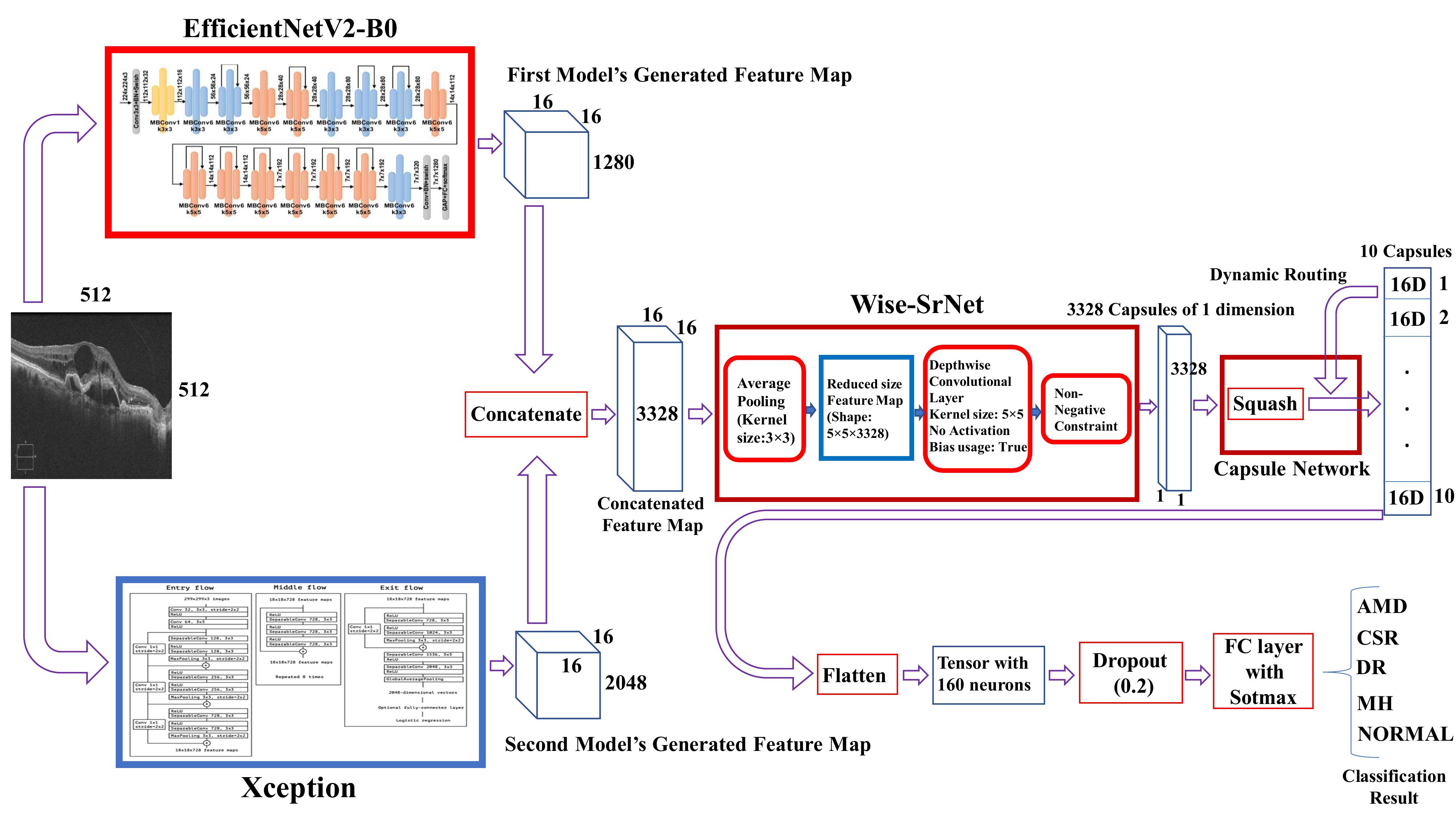}
\caption{This figure shows the architecture of our proposed model. The base feature extraction part works with concatenating Xception and EfficientNetV2-B0 models. Then the concatenated feature map will be fed to the post-architecture layers for improving the spatial resolution learning. The displayed classes at the end of the model are based on the OCTID dataset, which contains five classes. For the Kermany dataset, the number of classes will be four.}
\label{ensemble}
\end{figure*}

The architecture of the final ensemble model is presented in fig. \ref{ensemble}. We have trained it in an end-to-end manner. Utilizing the feature extraction strategies of these two models will make the extracted features richer due to the employment of multi-resolution features. This will also result in extracting more valuable features from the challenging parts of OCT images, as they contain tiny import regions that can reveal the disease signs. In other words, this ensemble model can resolve the possible failures caused through a single model (like waiving some information) by adopting the features of the other model. 

\subsection{Spatial Resolution Learning}
\label{23}

Spatial resolution is a very considerable matter in medical image analysis. Unlike other datasets, the hotspots of medical images that play an essential role in determining diseases may be small, so they can easily vanish when forwarding through compression layers (e.g., Global Average Pooling) at the end of the classification models. To fix this problem, we have designed a post-architecture model to prevent losing spatial data. Our post-architecture is made of two methods: Wise-SrNet \cite{rahimzadeh2021wise}, and Capsule networks \cite{sabour2017dynamic}.

\subsubsection{Wise-SrNet}
\label{231}

Wise-SrNet \cite{rahimzadeh2021wise} is a recently proposed architecture that solves the problem of losing spatial resolution data caused by usual compression and classification layers such as Global Average Pooling or Fully-connected layers. Unlike many other methods, this architecture will keep the spatial data while not increasing computation costs even on large images and is optimized for challenging conditions. 

We have applied the Wise-SrNet architecture at the end of our ensemble model (as it is depicted in fig. \ref{ensemble}) to compress the features without losing spatial resolution.

\subsubsection{Capsule Networks}
\label{232}

Capsule networks with dynamic routing algorithm \cite{sabour2017dynamic} has been published in 2017 and attracted the attention of many researchers. 
Capsules present a group of neurons whose activity vectors describe instantiation parameters of a specific entity, and their length shows the possibility of the existence of that entity. Capsule layers can make better interpretation from spatial relations of pixels than the convolutional layers.

Using the pooling layers in convolutional neural networks results in losing information, so the capsule layers replaced pooling layers with another method called routing by agreement. Routing by agreement method increases the impact of the capsules, which are more compatible with their capsule parents.

It is usual to use the squash function in capsule networks instead of usual activation functions like ReLU. The squash function normalizes the magnitude of vectors, and its output guides the model to rout between various training capsules.

\subsection{Our Architecture}
\label{24}

Our final architecture is presented in fig. \ref{ensemble}. The ensemble model, which was described in \ref{222} has been created, and the Wise-SrNet \cite{rahimzadeh2021wise} architecture was implemented on its end. The output of the Wise-SrNet is a compressed feature map with a 1x1x3328 shape that contains both the spatial and channel data. Our new architecture takes the final feature map and feeds it to a capsule network with some specific parameters to finally produce 10 number of 16D capsules using three routing iterations and the squash function. 

The main difference of our model with the typical capsule networks is that we did not use any reshaping, and the final feature map has entered the capsule network with shape 1x3328, meaning it is like to have 3328 number of 1D capsules. The second main difference is that the final output of our capsule network will be 10 number of 16D capsules while we have four classes in the first dataset and five classes in the second dataset. But why 10 capsules?

Our experiments prove that getting the final classification array from the capsule network will not work fine in our developed architecture. Hence, we gather a 10x16 matrix from the capsule network, which is an additionally processed version of the compressed feature map produced by Wise-SrNet. This way, we will maximize the spatial resolution learning process and increase its efficiency to get the higher possible results.
It must also be noted that the capsule networks' reconstruction process has not been utilized in our architecture and training procedure.

Then the capsule network's generated array (additionally compressed feature map) would be flattened and fed to a dropout and fully connected layer with softmax activation function to form the final classification array. Using dropout and an additional process of the capsule output array will also enhance the learning performance and decrease overfitting possibility.

In the next sections, we will show that our architrave certainly overcomes other models.

\section{Results and Comparison}
\label{3}

We implemented the models using Keras library \cite{chollet2015keras} on the Tensorflow backend. The training procedure was performed using Tesla P100 GPU and 12 GB RAM, provided by the Google Colab cloud.

For training our models on both datasets, all the images were resized to 512x512 pixels. We did not minimize the images so much because, in very small images, the tiny important areas of medical images may be lost.

The models have been trained using SGD optimizer with 0.9 momentum, and an adaptive learning rate starting with 0.045 initial value and 0.94 decay rate every two epochs (inspired by training method of \cite{chollet2017xception}). The batch size for our model was set to 10, and for other models, we used 15 batches. Categorical Cross Entroy was used as the loss function. Data augmentation techniques such as horizontal and vertical flipping, 10\% zoom range, 10\% width and height shifting, and 360-degree rotation were applied for enhancing the learning process and reduce overfitting. The data augmentations values were chosen somehow that the essential information of OCT images would not be lost.

The results of each dataset will be expressed separately in the next subsections. We have reported overall accuracy, mean sensitivity, and mean specificity metrics for evaluating the trained models. Overall accuracy and mean sensitivity are the same parameters and are defined by equation \ref{oa}.
Specificity is defined in equation. \ref{sp} and the mean-specificity is the weighted average of specificity between classes (Equation. \ref{msp}).

\begin{equation}
Overall\ accuracy\ (Mean\ Sensitivity) = \frac{TP}{TP+FN}
\label{oa}
\end{equation}

\begin{equation}
Specificity\ for\ each\ class= \frac{CTN}{CTN+CFP}
\label{sp}
\end{equation}

\begin{equation}
Mean\ Specificity= \sum_{C}^{} S_{C}N_{C}
\label{msp}
\end{equation}

In the deified equations, TP, FP, FN are the true positive, false positive, and false negative values of all the classes, and CTN, CFP refers to the true negative and false positive values of each class. In equation. \ref{msp}, S, N, and C refer to the specificity value of each class, number of images of each class, and number of classes, respectively.

\subsection{Experimental Results on the Kermany Dataset}
\label{31}

The datasets have been entirely described in \ref{21}.
For running our models on the Kermany dataset, all the models were initialized using transfer learning from the pre-trained weights on the ImageNet \cite{5206848} dataset. 

Although using transfer learning from the weights of the ImageNet dataset will not affect the learning efficiency of OCT images remarkably, it helps the model better extract image details like shape, edges, etc. So it can help to increase convergence speed compared to initializing the weights from scratch.

We continued training the models for 56 epochs. In this phase, EfficientNetV2-B0, Xception, and the ensemble models all enhanced with Wise-SrNet have been investigated and compared to our model.  As the effect of Wise-SrNet \cite{rahimzadeh2021wise} is clear on improving the classification results, we applied it also on other models to show that our architecture will suppress even the improved version of them. 

\begin{figure*}
\centering
\subfloat[Validation accuracy progress on the Kermany dataset]{\label{dataset1-fig}\includegraphics[width=0.5\linewidth]{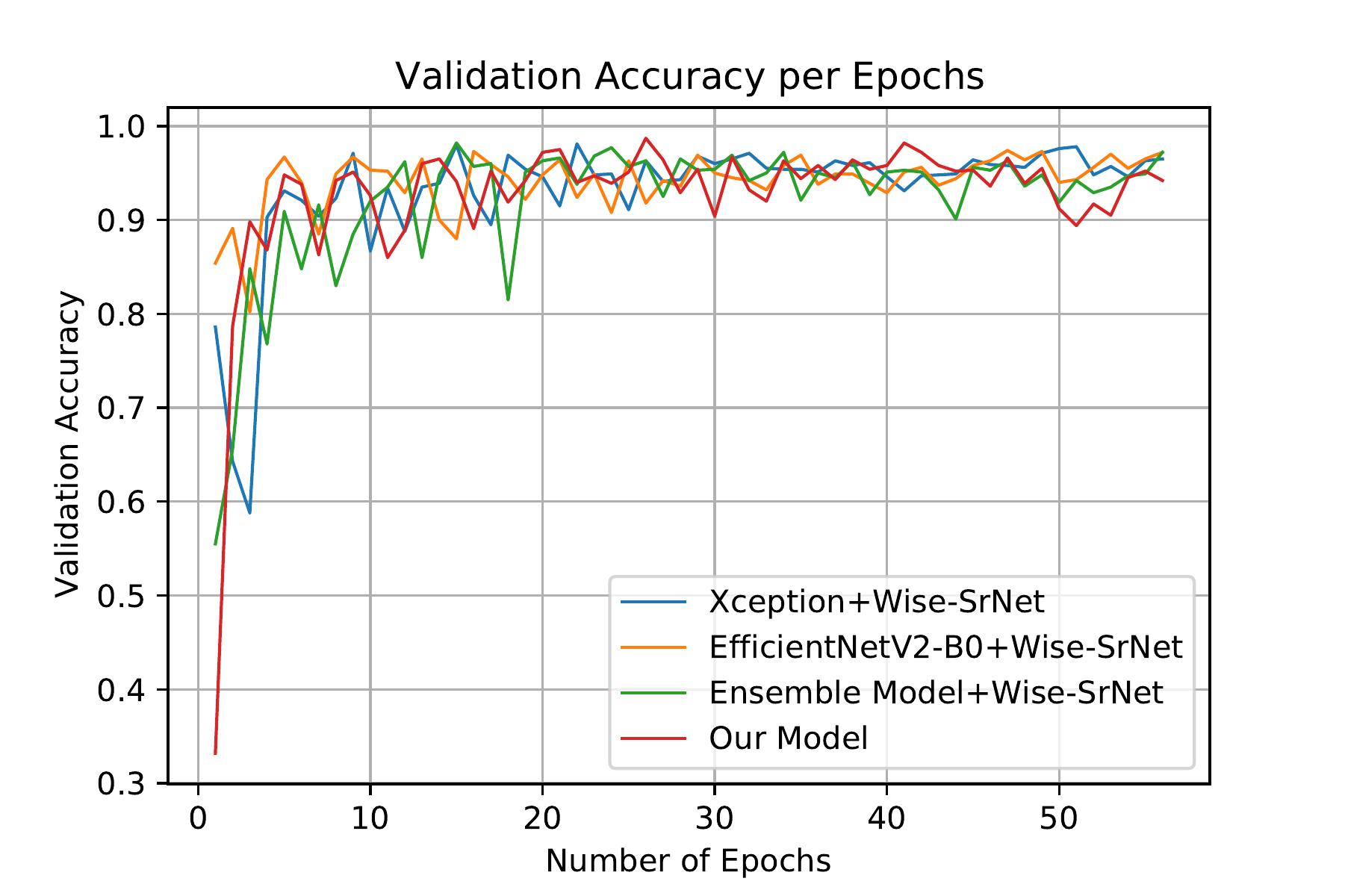}}
\subfloat[Validation accuracy progress on the OCTID dataset]{\label{dataset2-fig}\includegraphics[width=0.5\linewidth]{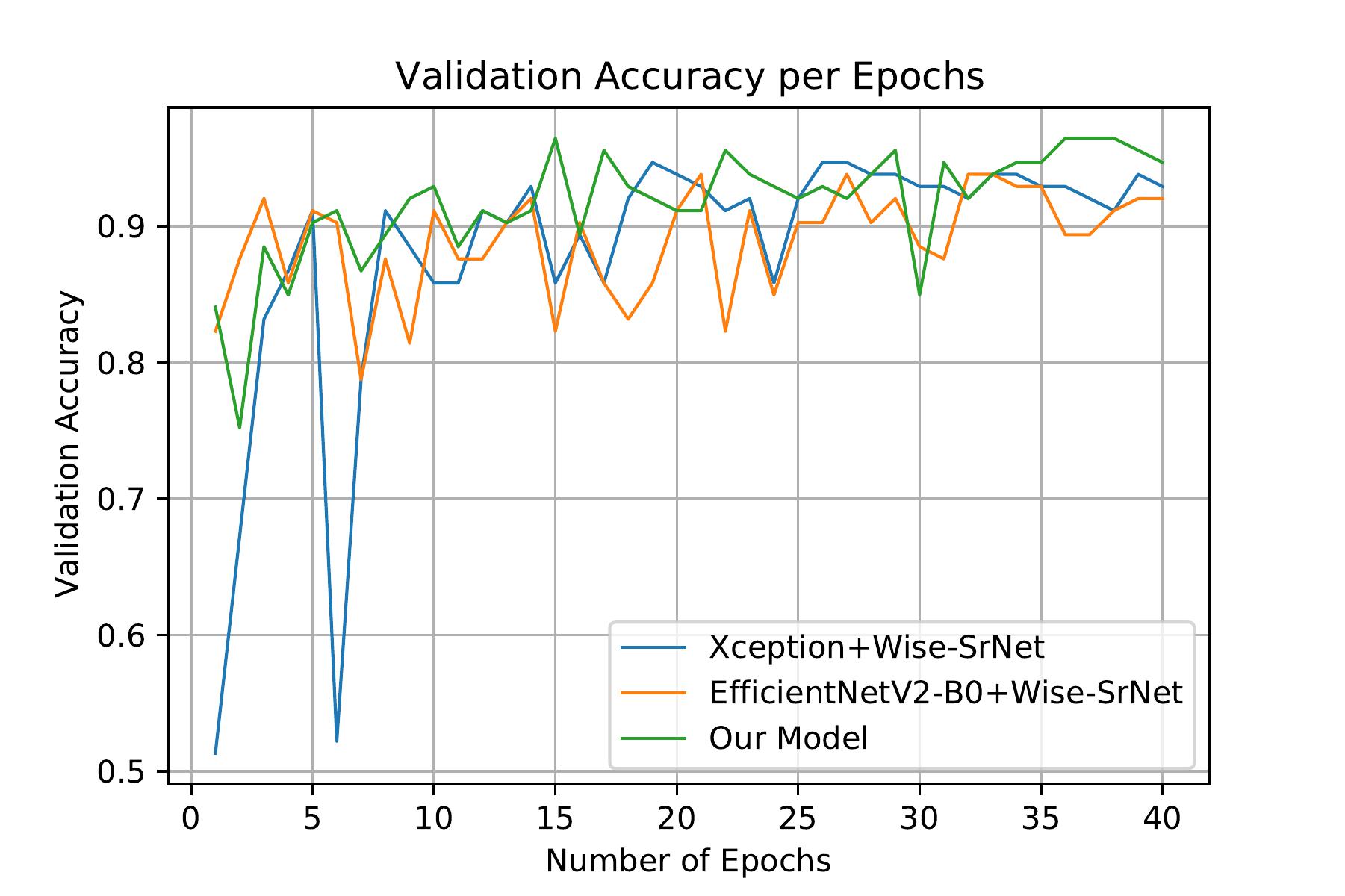}}
\caption{These figures show the validation accuracy per each epoch of training. The left and right figures are based on training on the Kermany, and OCTID datasets, respectively.}
\label{val_accuracy}
\end{figure*}

The results of this stage of our work are presented in table. \ref{dataset1-table}, and fig. \ref{dataset1-fig}, which confirms that our model shows higher performance. The accuracy and cross-entropy loss for each model has been mentioned in table. \ref{dataset1-table}. It must be noted that these metrics are evaluated on the test set.

\begin{figure*}[!ht]
\centering
\subfloat[Confusion matrix of our model on the Kermany dataset.
]{\label{confusion-data1}\includegraphics[width=0.5\linewidth]{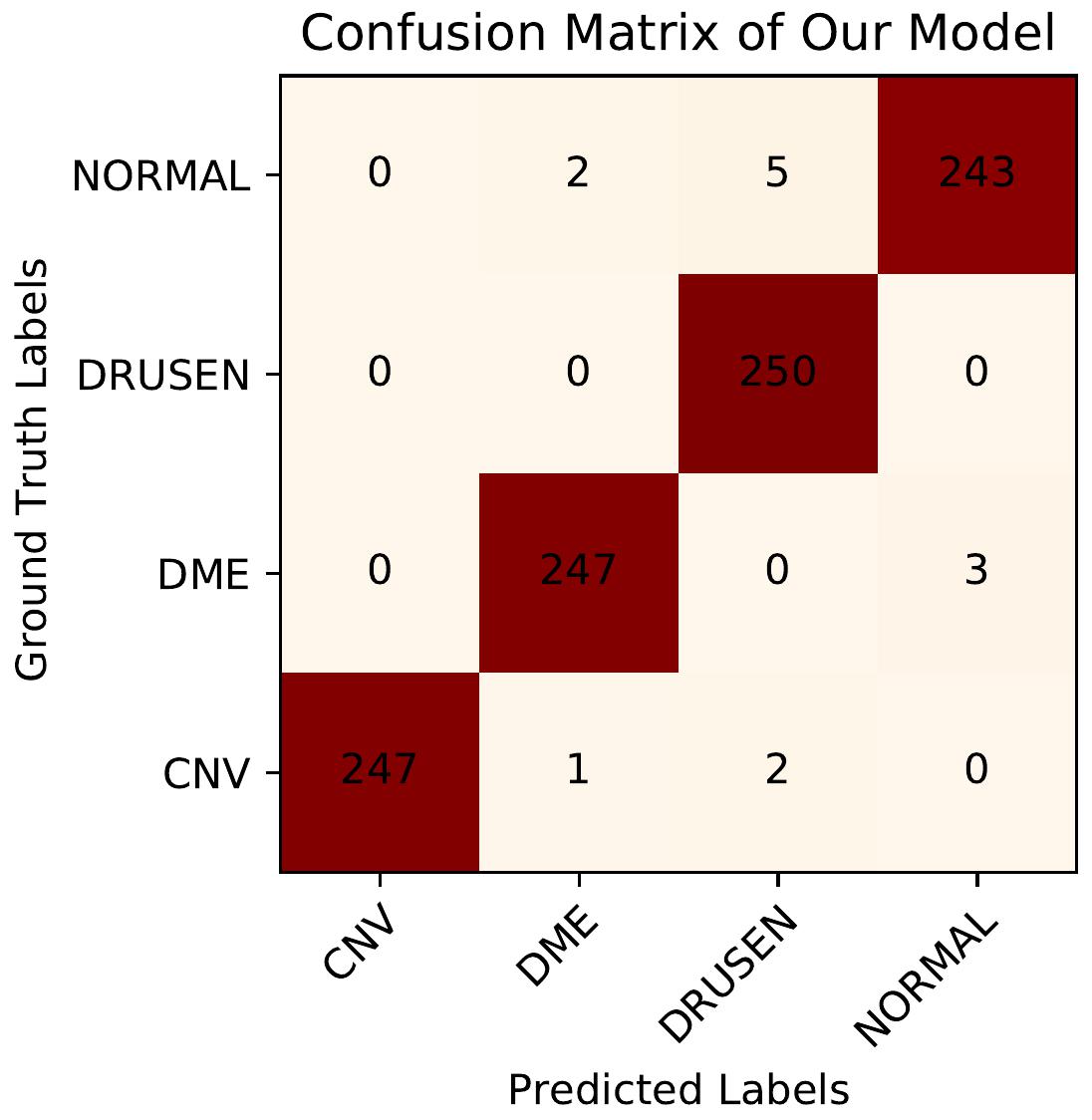}}
\subfloat[Confusion matrix of our model on the OCTID dataset.]{\label{confusion-data2}\includegraphics[width=0.5\linewidth]{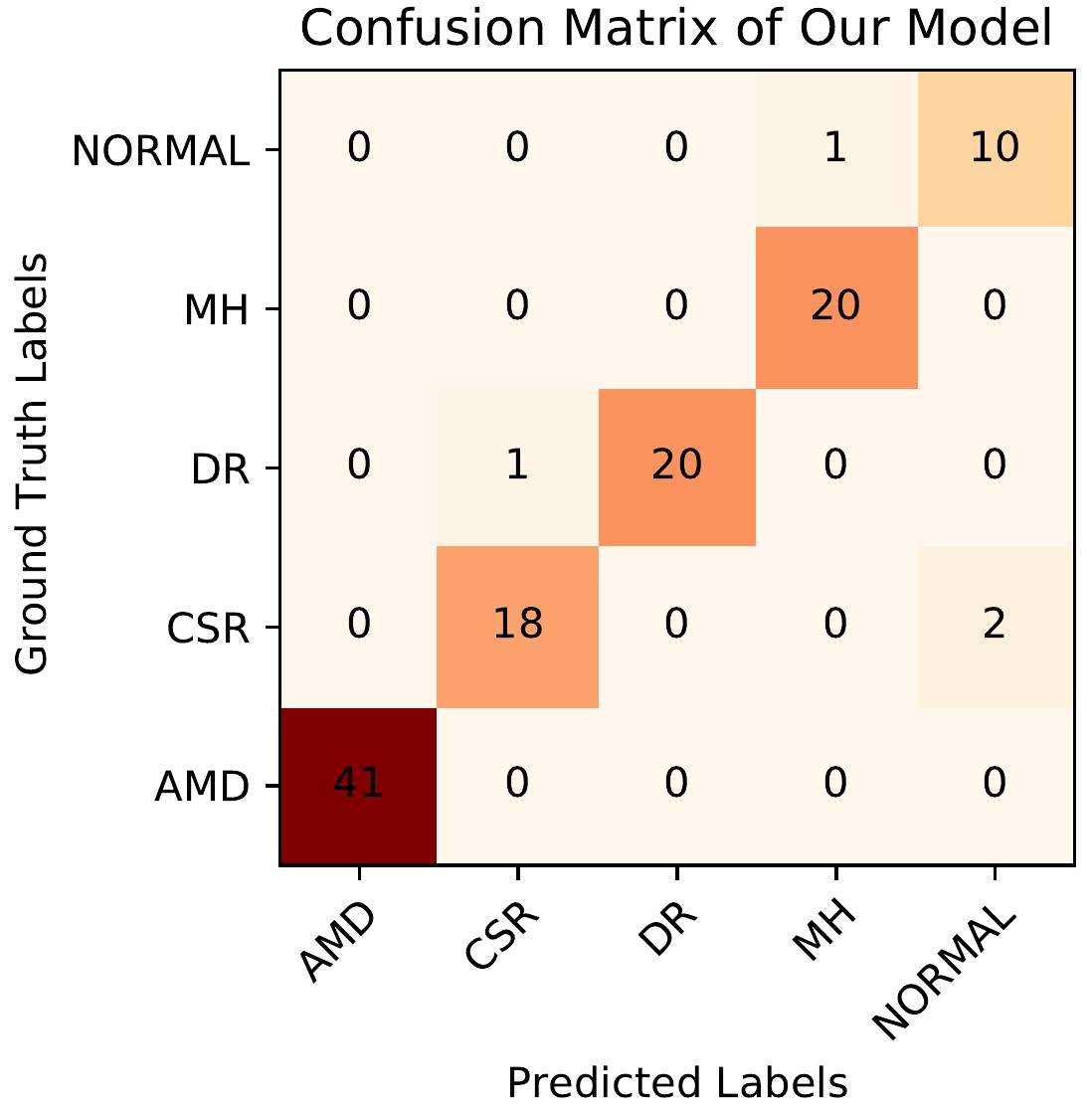}}
\caption{The confusion matrices of our model are represented in this figure.}
\label{confusion}
\end{figure*}

The confusion matrix of our model on this dataset is also presented in fig. \ref{confusion-data1}.

\begin{table}
\caption{The evaluation results of our experiments on the Kemrnay dataset are reported in this table.}
\label{dataset1-table}
\centering
\large
\begin{adjustbox}{width=1\linewidth}
\begin{tabular}{|c|c|c|}
\hline
Model                        & \begin{tabular}[c]{@{}c@{}}Mean\\ Sensitivity\end{tabular} & \begin{tabular}[c]{@{}c@{}}Mean\\ Specificity\end{tabular} \\ \hline
Xception+Wise-SrNet          & 0.9810                                                     & 0.9937                                                     \\ \hline
EfficientNetV2-B0+Wise-SrNet & 0.9740                                                     & 0.9913                                                     \\ \hline
Ensemble Model+Wise-SrNet    & 0.9820                                                     & 0.9940                                                     \\ \hline
Our Model                    & \textbf{0.9870}                                            & \textbf{0.9957}                                            \\ \hline
\end{tabular}
\end{adjustbox}
\end{table}

\subsection{Experimental Results on the OCTID Dataset}
\label{32}

In this stage, we transferred the weights of our models trained on the Kermany dataset (that includes many images) in the previous section to initialize the models' weights for training on the OCTID dataset.

Although most of the classes between these two datasets are different, as both datasets include OCT images, their attributes are similar. Therefore, using transfer learning from a well-trained model will significantly impact increasing training speed and efficiency, especially in challenging conditions like having few training images.

Three models were selected for investigation on this datasets: Xception with Wise-SrNet, EfficientNetV2-B0 with Wise-SrNet, and Our model. We trained them for 40 epochs.

Table. \ref{dataset2-table}, and fig. \ref{dataset2-fig} displays the obtained results of the trained models. Comparing the results clarify that our model performs more accurately.
The confusion matrix of our model on this dataset is also presented in fig. \ref{confusion-data2}.

\begin{table}
\caption{The evaluation results of our experiments on the OCTID dataset are reported in this table.}
\label{dataset2-table}
\centering
\large
\begin{adjustbox}{width=1\linewidth}
\begin{tabular}{|c|c|c|}
\hline
Model                        & \begin{tabular}[c]{@{}c@{}}Mean\\ Sensitivity\end{tabular} & \begin{tabular}[c]{@{}c@{}}Mean\\ Specificity\end{tabular} \\ \hline
Xception+Wise-SrNet          & 0.9469                                                     & 0.9883                                                     \\ \hline
EfficientNetV2-B0+Wise-SrNet & 0.9381                                                     & 0.9883                                                     \\ \hline
Our Model                    & \textbf{0.9646}                                            & \textbf{0.9943}                                            \\ \hline
\end{tabular}
\end{adjustbox}
\end{table}

\subsection{Comparison with Other Works}
\label{33}

Most of the previous works in this field are based on detecting one or two diseases from OCT images \cite{thomas2021automated, kaymak2018automated, hassan2021deep, thomas2021automated}. As we explore detecting several diseases in two different datasets, there are few works whose research criteria are similar to ours. 
In table. \ref{comparison}, we have compared our model with some other researches.

\begin{table}
\caption{Comparing our model on two different datasets with some other methods.}
\label{comparison}
\centering
\large
\begin{adjustbox}{width=1\linewidth}
\begin{tabular}{|l|l|l|l|l|}
\hline
Methods                             & Dataset                          & \begin{tabular}[c]{@{}l@{}}Training\\  Images\end{tabular} & \begin{tabular}[c]{@{}l@{}}Evaluation\\  Images\end{tabular} & \multicolumn{1}{c|}{\begin{tabular}[c]{@{}c@{}}Overall\\ Accuracy\end{tabular}} \\ \hline
Singh \cite{singh2021uncertainty}      & OCTID  & 459                                                        & 113                                                          & 0.9140                                                                          \\ \hline
Our Model                           & OCTID  & 459                                                        & 113                                                          & \textbf{0.9646}                                                                 \\ \hline
IncptionV3 \cite{kermany2018identifying} & Kermany  & 108,312                                                    & 1000                                                         & 0.9610                                                                          \\ \hline
Our Model                           & Kermany  & 3213                                                       & 1000                                                         & \textbf{0.9870}                                                                 \\ \hline
\end{tabular}
\end{adjustbox}
\end{table}

It is evident from table. \ref{comparison} that our model outperforms the previous methods. The notable point is that on the Kermany dataset, we only used 3213 training images while the original paper \cite{kermany2018identifying} adopted the whole 108,312 images for training the models. It can be concluded that our proposed model can reach higher accuracy when training on much fewer data. This statement shows the learning capability of our model and introduced methods.

\section{Discussion}
\label{4}

The reported results in section \ref{3} illuminate that our model outperforms other models by utilizing multi-resolution features extracted by two robust deep convolutional networks and having the privilege of employing a post-architecture model for improving spatial resolution learning.

Experiments on the Kermany dataset show that our model increases the classification accuracy by 2.6\% than the InceptionV3 model \cite{kermany2018identifying} while using less than 3\% of training data of the InceptionV3 model. This result shows the high performance and capability of our model. 
Our model also achieved higher results compared to its component models. It achieved 98.70\% overall accuracy on 1000 test images of the Kermany dataset.

On the OCTID dataset, investigations show that our model achieved higher overall accuracy within 1.77\% to 5.06\%. Furthermore, besides the architecture of our proposed model, it is noteworthy that using transfer learning from pre-trained weights on OCT images has increased classification accuracy very well.
 Our model achieved 96.46\% overall accuracy on 113 test images, meaning it correctly classified 109 images out of 133 images.

Researchers can inspire from the methods we used in our model and create more accurate models in the next research works also.

\section{Conclusion}
\label{5}
This paper presented a novel Computer-Aided Diagnosis (CAD) system for detecting several crucial diseases from optical coherence tomography (OCT) images. CAD systems can be great assistants for ophthalmologists in precise and early retinal pathology. Our proposed ensemble model is capable of extracting high-quality and multi-resolution features that can represent different aspects of OCT images. Our model generates these features by adopting the feature extraction strategies of two concatenated convolutional neural networks.
In the next stage, as the spatial resolution is a notable factor in medical images, we introduced a new post-architecture model for enhancing spatial resolution learning. This post-architecture can be applied to any feature extraction model and enhances the capability of extracting richer spatial data from the final generated feature map by employing Wise-SrNet and capsule networks.
We have gathered two of the largest open-source datasets, including OCT images from seven classes: AMD, CSR, DR. MH, CNV, DME, DRUSEN, and Normal cases. We ran our experiments on these two datasets and compared the achieved results of our model with some other methods and neural networks. Our investigations showed that our model achieves up to 5\%  more accuracy compared to other CAD methods.
The code of this paper is shared at \href{https://github.com/mr7495/OCT-classification}{https://github.com/mr7495/OCT-classification}.

\section*{Acknowledgment}

This work was supported by the \href{https://faraai.ir}{FaraAI} (Faraz Hoosh Sharif) company.
The authors wish to sincerely thank Dr.Hesam Joghtaee for his guidance
and valuable comments.

This is a preprint of an article published in the ICCKE 2021 conference. The final authenticated version is
available online at \href{https://doi.org/10.1109/ICCKE54056.2021.9721471}{10.1109/ICCKE54056.2021.9721471}.

\bibliographystyle{IEEEtran}
\bibliography{IEEE}

\end{document}